\documentclass[11pt]{article}
\vspace{7cm} \textwidth 14cm \textheight 21.6cm
\usepackage{amsmath}
\usepackage{amsfonts}
\usepackage{authblk}
\usepackage{url}
\usepackage[english]{babel}
\usepackage[dvipsnames]{xcolor}
\usepackage{hyperref}
\usepackage{float}
\usepackage[pdftex]{graphicx}
\usepackage{caption}
\usepackage{subcaption}
\usepackage[utf8]{inputenc}
\usepackage{booktabs}
\usepackage{changes}
\usepackage{latexsym,amsthm,amscd}
\usepackage{booktabs}
\usepackage[margin=1in]{geometry}
\usepackage{longtable}
\newcounter{alphthm}
\usepackage{tikz}
\usetikzlibrary{shapes.geometric, arrows}
\tikzstyle{startstop} = [rectangle, rounded corners, minimum width=1cm, minimum height=1cm,text centered, draw=black, fill=blue!30]
\tikzstyle{io} = [trapezium, trapezium left angle=70, trapezium right angle=110, minimum width=3cm, minimum height=1cm, text centered, draw=black, fill=blue!30]
\tikzstyle{process} = [rectangle, minimum width=5cm, minimum height=1cm, text centered, draw=black, fill=white!30]
\tikzstyle{decision} = [ellipse, minimum width=3cm, minimum height=1cm, text centered, draw=black, fill=white!30]
\tikzstyle{arrow} = [thick,->,>=stealth]
\tikzstyle{edge from parent}=[draw,dashed,thick,->,>=stealth,black]
\tikzstyle{process} = [rectangle, minimum width=3cm, minimum height=1cm, text centered, text width=3cm, draw=black, fill=white!30]
\theoremstyle{remark}
\usepackage{amssymb, graphicx, tcolorbox, wasysym,  pifont}
\numberwithin{equation}{section}

\newtheorem*{theorem*}{Theorem}

\theoremstyle{definition}

\theoremstyle{remark}

\title{Wildfire Propagation Modeling using Satellite-Derived Parameters and Generalized Elliptical Frames }
\date{\today}

\author[1]{Hengameh R. Dehkordi}
\affil{Center of Mathematics, Computing, and Cognition, Federal University of ABC, 09210-580, Santo André, SP, Brazil\\
\texttt{hengameh.r@ufabc.edu.br}\\
}

\begin{document}
\maketitle

\begin{abstract}



Wildfires pose significant threats to ecosystems and communities, yet accurately modeling fire spread remains challenging, particularly in regions where environmental and fuel data are scarce or unavailable. This study introduces an innovative conceptual and methodological framework for simulating wildfire propagation and estimating the rate of spread using a hybrid geometric and data-driven approach that relies exclusively on multi-source satellite observations. By removing the dependence on detailed fuel models or extensive in situ terrain measurements, the proposed methodology offers a practical and scalable alternative for regions with limited ground-based information.

The framework integrates thermal fire-front detections, atmospheric conditions, and vegetation indices using two complementary geometric modeling strategies. The first strategy applies the Huygens principle, where generalized elliptical frames are expanded locally at every point along the fire perimeter, and their combined envelope forms the evolving wavefront. This method is best suited for situations in which environmental variables—such as wind fields or live fuel moisture—are available and can be incorporated to refine the anisotropic spread function. The second strategy relies solely on the generalized elliptical frames themselves: for each time step, an elliptical frame is constructed from the inferred head and back rates of spread and wind, and the burned area is obtained by enclosing the region determined by these curves. Together, these two methods provide a flexible toolkit that adapts to both data-rich and data-limited conditions while retaining a unified geometric interpretation of wildfire spread.

To demonstrate the applicability of the method, we present a case study based on the Eaton Fire, January 2025, using publicly available multi-day satellite imagery. Despite the absence of complete operational datasets for that event, the model—driven only by satellite-derived parameters—reproduces key qualitative features of the observed propagation pattern, underscoring the flexibility and robustness of the proposed approach in data-limited contexts.

Conceptually clear, computationally efficient, and highly adaptable, this methodology bridges advanced geometric modeling with modern satellite-based fire monitoring. The framework establishes a foundational structure for future developments and supports applications ranging from local operational fire management to continental-scale environmental assessment.

\end{abstract}


\section{Introduction}
Fire propagation is a highly complex phenomenon that requires sophisticated modeling to predict its behavior accurately. Effective fire propagation models must integrate thermal, atmospheric, vegetation data, and terrain topography to simulate how a fire spreads over time and across different terrains~\cite{andrews2018rothermel}. Factors such as wind speed and direction, temperature, fuel type, fuel moisture, and topography play crucial roles in fire dynamics, and models considering these variables tend to yield more reliable predictions~\cite{andrews2018rothermel}. By combining diverse satellite sources—including thermal and atmospheric data—with recent advances in mathematics \cite{dehkordi2022applications, dehkordi2024approximated}, this work aims to offer a comprehensive understanding of wildfire behavior, ultimately enhancing fire prediction and management strategies.

Satellite data used in fire propagation modeling can be categorized into three main types: atmospheric (e.g., wind, temperature), thermal (e.g., active fire detection), and vegetation (e.g., fuel type, dryness)~\cite{abdelkhalek2019simulation}. Wind data are critical, as strong winds can carry embers over large distances, igniting new fires ahead of the main front. Satellite instruments like Moderate Resolution Imaging Spectroradiometer (MODIS), Visible Infrared Imaging Radiometer Suite (VIIRS), and Sea and Land Surface Temperature Radiometer (SLSTR) measure brightness temperature and detect hotspots, providing real-time data on fire intensity and spread, often sourced from organizations like European Organization for the Exploitation of Meteorological Satellites and National Aeronautics and Space Administration (NASA)~\cite{ghali2023deep, martins2025analysing}.

Thermal sensors provide important information about active fire fronts and fire radiative power, enabling the estimation of directional rates of spread~\cite{abdelkhalek2019simulation}. Atmospheric data, such as wind speed and direction, are used to model anisotropic fire behavior, capturing how wind influences fire spread in different directions~\cite{fu2023satellite}.

Vegetation indices derived from optical satellites assess fuel conditions—such as biomass availability and moisture content—allowing for spatial refinement of the fire spread model~\cite{ling2025research}. 
Vegetation data are essential, as fuel types and their spatial distribution greatly influence fire dynamics~\cite{di2025global}. Satellite imagery can identify different vegetation classes and assess vegetation health—using indices like the Normalized Difference Vegetation Index (NDVI)—and moisture levels, with dry, combustible vegetation favoring rapid fire spread~\cite{up42fire}. These datasets allow for detailed mapping of fuel conditions, which are vital for accurate modeling~\cite{hernandez2006fire}.

The novelty of this work lies in its integrative framework, which combines multi-source satellite data—thermal, atmospheric, and vegetation—with a geometric formulation grounded in Huygens’ principle and the generalized elliptical frames to construct a comprehensive model of wildfire propagation. Rather than relying on computationally intensive simulations, the proposed method employs a generalized elliptical representation to capture the anisotropic spread of fire. This formulation provides a mathematically tractable yet physically meaningful description of fire front evolution, offering a balance between model complexity and practical applicability.

 Our technique begins with thermal satellite time series to estimate the initial fire front, followed by numerical refinements informed by vegetation characteristics and wind conditions. A key contribution of this study is the introduction of a correction function that adjusts thermal-derived spread rates using vegetation indices and environmental variables. The generalized elliptical formulation enables consistent calibration of anisotropic fire spread models across spatial and temporal scales, effectively bridging physics-based and data-driven approaches while improving predictive accuracy and operational applicability.  

Furthermore, the proposed hybrid framework integrates real-time fire detections with remote sensing information on vegetation, allowing the modeling of wildfire propagation from local to regional scales. By accounting for spatial variability in fuel properties, the correction-based method refines the anisotropic spread functions, ensuring that simulated fire dynamics align more closely with observed behavior.

It is important to note that this study is intentionally focused on developing the theoretical and computational foundations of the proposed framework rather than performing case-specific validation. The absence of real fire event comparisons or numerical calibration is a deliberate choice, as the present goal is to construct a generalizable and conceptually robust modeling structure. This foundational stage is essential to ensure that the framework can later be adapted and validated across different regions and vegetation types. Future work will extend this approach to real satellite time series and meteorological datasets to evaluate its predictive performance under operational conditions.

The remainder of this study is organized as follows. Section 2 presents the preliminary concepts. Section 3 introduces our strategy for estimating and refining the rate of spread, accompanied by illustrative examples that clarify the proposed methodology. Finally, Section 4 concludes the paper with a summary of the main findings and outlines future research directions.

\section{Satellite Remote Sensing: Platforms, Thermal Products, and Vegetation Indices for Wildfire Monitoring}

In this section, we provide a concise overview of the satellite datasets and products employed to estimate wildfire rates of spread, ensuring the paper is self-contained.

The estimation combines thermal, wind, and vegetation information from multiple satellite platforms. Thermal imagery from geostationary satellites captures fire hotspots and thermal anomalies with high temporal frequency, enabling the monitoring of rapid fire front dynamics. Polar-orbiting sensors complement this with high-resolution observations of fire perimeters and vegetation conditions, allowing for precise identification of fire front positions and their temporal evolution.

Atmospheric data provide wind speed and direction, while vegetation indices inform on fuel moisture and availability. These variables are integrated into numerical methods to predict head fire and backfire rates of spread under varying fuel and wind conditions, forming the basis for a comprehensive, satellite-driven wildfire propagation model.
 
\subsection{Thermal Infrared Remote Sensing}

Thermal infrared (TIR) remote sensing is fundamental to satellite-based wildfire monitoring, enabling the detection of active fires by pinpointing areas with abnormally high surface temperatures. This capability ensures dependable fire detection even under challenging conditions, such as dense smoke or nighttime observations. Satellites like MODIS and VIIRS provide global, near-real-time active fire products, such as MOD14, which identify fire hotspots and assist in the rapid assessment of fire extent and location~\cite{zheng2023research}.

Beyond simple detection, TIR data also facilitate the retrieval of Fire Radiative Power (FRP), which measures the rate of radiant energy release from burning areas and serves as an indicator of fire intensity and combustion rate. FRP products derived from instruments aboard VIIRS and the Geostationary Operational Environmental Satellites (GOES) are essential for modeling fire behavior, fuel consumption, and emissions. While GOES offers high-frequency observations every 5--15 minutes, the polar-orbiting VIIRS sensor provides 2--4 overpasses per day, together enabling near-continuous tracking of fire dynamics over time~\cite{li2018comparison, schroeder2014viirs}.

Beyond active fire detection, TIR observations are instrumental in retrieving Land Surface Temperature (LST) over areas not directly affected by fires. LST represents the radiometric temperature of the Earth's surface and is a key indicator of the surface energy balance, soil moisture status, and vegetation stress. This capability enables diverse environmental applications, including drought stress monitoring, pre-fire vegetation assessment, and post-fire heat retention analysis. TIR sensors aboard satellites such as MODIS, VIIRS, and Landsat's Operational Land Imager (OLI)/TIRS measure upwelling thermal radiation, which—after appropriate atmospheric correction—is converted into surface temperature. Consequently, LST provides valuable insights into vegetation stress and surface drying patterns that often precede wildfire ignition.

Table~\ref{tabsat} summarizes the main satellite missions and sensors commonly used for wildfire detection and monitoring, along with their temporal and spatial resolutions~\cite{justice2002modis, wagenbrenner2016downscaling}.

\begin{center}
\small
\begin{longtable}{@{}llll@{}}
\caption{Satellites frequently employed for wildfire monitoring.}\label{tabsat} \\
\toprule
\textbf{Satellite} & \textbf{Instrument} & \textbf{Overpass Frequency} & \textbf{Resolution} \\
\midrule
\endhead
\textbf{Terra}       & MODIS     & $\sim$10:30 AM LT         & 1 km    \\
\textbf{Aqua}        & MODIS     & $\sim$1:30 PM LT          & 1 km    \\
\textbf{Suomi-NPP}   & VIIRS     & $\sim$1:30 AM/PM          & 375 m   \\
\textbf{NOAA-20}     & VIIRS     & $\sim$1:30 AM/PM          & 375 m   \\
\textbf{GOES-16}     & ABI       & Every 5--15 minutes       & 2 km    \\
\textbf{GOES-17}     & ABI       & Every 5--15 minutes       & 2 km    \\
\textbf{Sentinel-3}  & SLSTR     & 1--2 times per day        & 1 km    \\
\textbf{MSG-SEVIRI}  & SEVIRI    & Every 15 minutes          & 3 km    \\
\bottomrule
\end{longtable}
\end{center}

In Table~\ref{tabsat}, Suomi-NPP refers to the \textit{Suomi National Polar-orbiting Partnership}, NOAA denotes the \textit{National Oceanic and Atmospheric Administration}, and MSG-SEVIRI corresponds to the \textit{Meteosat Second Generation} satellites equipped with the \textit{Spinning Enhanced Visible and InfraRed Imager}  \cite{justice2002modis, wagenbrenner2016downscaling}.

\subsection{Live Fuel Moisture Content }

LFMC, defined as the ratio of water mass to dry biomass in live vegetation, is a key biophysical parameter influencing both wildfire ignition probability and the rate of fire spread~\cite{Yebra2013}. It reflects the physiological condition of vegetation and its resistance to combustion. Due to the spatial and temporal limitations of field sampling, satellite-based LFMC estimation provides an effective alternative for assessing fire danger over large regions~\cite{rao2020sar}. In this study, LFMC is employed as one of the principal variables used to predict the head fire and backfire rates of spread.

Empirical and semi-empirical models commonly relate LFMC to spectral vegetation indices that are sensitive to vegetation water content~\cite{Yebra2013}. When auxiliary data such as surface temperature or temporal information are unavailable, a single vegetation index—typically NDVI—can be used as a simplified proxy for LFMC, since greener and more vigorous vegetation generally contains higher water content~\cite{Chuvieco2004, chuvieco2004combining}. A widely used linear relationship is expressed as
\[
{LFMC} = \mathfrak{a} \cdot VI + \mathfrak{b},
\]
where $VI$ denotes the vegetation index, and $\mathfrak{a}$ and $\mathfrak{b}$ are regression coefficients determined from in place measurements and satellite observations~\cite{Yebra2013, rao2020sar}.

Nevertheless, \cite{Yebra2013} highlights that no single spectral index can reliably estimate LFMC across all vegetation types, especially in dense canopies where NDVI often saturates. They suggest combining complementary variables that capture different physical aspects of vegetation condition. Consequently, the most accurate empirical models incorporate: (i) greenness indices like NDVI, which represent vegetation activity; (ii) thermal variables (e.g., LST) that indicate surface energy balance and moisture stress; and (iii) temporal variables (e.g., Day of Year, DOY) that account for phenological variation, as illustrated in Table 1 of \cite{Yebra2013}.

\subsection{Vegetation Satellites Dataset}

NDVI is a prevalent remote sensing metric employed to assess vegetation greenness and health by evaluating the differential reflectance between the near-infrared (NIR) and red (RED) wavelengths captured by satellite sensors. Healthy vegetation is characterized by strong reflection of NIR light and significant absorption of red light due to chlorophyll, enabling NDVI to act as an indicator of photosynthetic activity \cite{hernandez2006fire}. 

Mathematically, the NDVI is expressed as 
$$NDVI = \frac{(NIR - RED)}{(NIR + RED)},$$
where NIR and RED denote the spectral reflectance in the near-infrared and red bands, respectively \cite{hernandez2006fire}. Because both NIR and RED can range from 0 to 1, NDVI values range from –1 to +1, with higher values (typically above 0.5) signifying dense, healthy vegetation, whereas lower or negative values indicate barren surfaces, water, or clouds \cite{glenn2019evaluation}.  NDVI is a crucial parameter for estimating the LFMC, which directly affects fire ignition and propagation dynamics \cite{nature2025, Yebra2013, rao2020sar}. Prior to a fire, NDVI detects the presence of fuel, and LST detects dryness. During a fire, thermal sensors monitor hotspots and intensity, and post-fire, NDVI and LST assess damage, regrowth, and residual heat.

Among vegetation indices, EVI (Enhanced Vegetation Index) offers a refined alternative to NDVI for assessing vegetation condition and dynamics. EVI improves upon NDVI by reducing atmospheric and canopy background noise and by enhancing sensitivity in high-biomass regions~\cite{qiu2018}. It maintains strong responsiveness over dense vegetation canopies and is widely applied in studies of ecosystem productivity, drought detection, and LFMC estimation~\cite{Yebra2013}.



Table~\ref{NDVI} summarizes the main satellite instruments commonly used to estimate  NDVI and EVI, along with their respective temporal and spatial resolutions (see~\cite{modisMOD13, firmsVIIRSFireHotspots, usgsLandsat8, copernicusSentinel2, eoportalCBERS4A}).

\begin{center}
\small
\begin{longtable}{@{}llll@{}}
\caption{Satellites frequently employed for  vegetation indices monitoring and condition analysis.}\label{NDVI}\\
\toprule
\textbf{Satellite} & \textbf{Instrument} & \textbf{Overpass Frequency} & \textbf{Resolution} \\
\midrule
\endhead
\textbf{Aqua}         & MODIS     & $\sim$1:30 PM LT          & 250 m -- 1 km \\
\textbf{NOAA-20}      & VIIRS     & $\sim$1:30 AM/PM          & 375 m \\
\textbf{Sentinel-2A}  & MSI       & 5-day revisit             & 10 m -- 20 m \\
\textbf{Landsat-8}    & OLI/TIRS  & 16-day revisit            & 30 m \\
\textbf{CBERS-4/4A}   & MUXCam/WPM/WFI & Varies by instrument (WFI: $\sim$5-day revisit) & 20 m -- 55 m \\
\bottomrule
\end{longtable}
\end{center}

\subsection{Estimation of Coefficients for LFMC}
Inspired by \cite{Chuvieco2004}, we use the empirical relationship to estimate LFMC from remote sensing data as below
\begin{equation}
    {LFMC} = A + B \cdot {NDVI} + C \cdot {LST} + D_1\cdot\sin\left(2\pi \frac{{DOY}}{365}\right) + D_2\cdot
\cos\left(2\pi \frac{{DOY}}{365}\right),
    \label{eq:lfmc_model}
\end{equation}
where $A$, $B$, $C$,  $D_1$ and $D_2$ are regression coefficients to be estimated and ${DOY}$ denotes the day of the year.  Two last terms naturally captures {seasonal variation} without artificial jumps at year-end. 

The empirical relationship~\ref{eq:lfmc_model}
combines spectral, thermal, and temporal variables to represent vegetation condition and its seasonal evolution. 
Each coefficient has a clear physical interpretation, as summarized below \cite{chuvieco2004combining}.

\begin{enumerate}
    \item {Intercept (${A}$):} Represents the baseline or reference LFMC value when all predictors are centered or zero. It ensures the estimated LFMC remains within a realistic range and accounts for the mean moisture condition of the study region.

    \item {NDVI (Coefficient ${B}$; Positive Correlation):} 
    Higher NDVI values correspond to denser and healthier vegetation with greater photosynthetic activity and typically higher water content. 
    Therefore, the coefficient ${B}$ is expected to be {positive}. 
    However, NDVI may saturate in dense or high-biomass canopies, reducing sensitivity to LFMC variations. 
    In such cases, it can be complemented or substituted by indices, such as EVI, that are more responsive to canopy water content.

    \item {LST (Coefficient ${C}$; Negative Correlation):} 
    LST is inversely related to vegetation moisture. 
    Higher LST values indicate increased thermal stress and reduced evaporative cooling, conditions typically associated with lower LFMC. 
    Thus, the coefficient ${C}$ is expected to be {negative}.

    \item {Seasonal Terms (Coefficients ${D_1}$ and ${D_2}$):} 
    The sine and cosine terms of DOY capture the seasonal cycle of vegetation moisture driven by phenology and climate. 
    Together, they allow the model to represent annual fluctuations in LFMC and to adjust for regional phase shifts in moisture dynamics. 
    Typically, ${D_1}$ corresponds to the amplitude of the wet or growing season (often {positive}), 
    while ${D_2}$ accounts for the phase shift and may take {positive or negative} values depending on the timing of maximum and minimum LFMC across regions.
\end{enumerate}

In the following, we describe practical strategies for estimating these coefficients under different data availability scenarios.

\paragraph{Data Reprocessing}

Prior to model calibration, all predictor variables are harmonized to a common spatial resolution and coordinate projection and the ${NDVI}$, ${LST}$, and ${DOY}$ indices are resampled to a consistent grid (typically 1~km or the native sensor resolution), and low-quality or cloudy pixels are removed and then we consider the following cases:
\begin{itemize}
\item{Case 1: Field-Based Calibration }

When field-measured LFMC data are available, the coefficients $A$, $B$, $C$,  $D_1$ and $D_2$ can be estimated directly from coincident satellite and field observations. That is, for each in-situ LFMC measurement, ${NDVI}$, ${LST}$, and ${DOY}$ are extracted from the nearest satellite overpass.

\item{Case 2: Absence of Field LFMC Data}

When in-situ LFMC measurements are not available, we adopt coefficients from published studies conducted in similar vegetation and climatic conditions, while acknowledging potential regional bias.

\item{Case 3: Regionalization of Model Coefficients}

For large or environmentally heterogeneous study areas, regionalization improves model reliability and transferability.
Two complementary strategies can be adopted:

\begin{itemize}
    \item \textbf{Stratified modeling:} subdivide the area by land-cover or vegetation class (e.g., forest, shrubland, grassland) and fit separate models for each stratum. 
    \item \textbf{Mixed-effects modeling:} estimate global fixed effects for NDVI and LST, while allowing regional random intercepts or slopes to capture local differences in vegetation response.
\end{itemize}

\end{itemize}

\paragraph{Implementation and Practical Considerations}

\color{black} We  use MATLAB or Python programming and  use a model, such as Ordinary Least Squares, to estimate  the coefficients. We can evaluate the model using  RMSE, MAE, and $R^2$ on independent validation data. Spatial cross-validation (e.g., leave-region-out) is applied to evaluate the generalization ability of each approach. 
When sample size per region is limited, partial pooling via hierarchical models or regularization is preferred over independent regional fits. \color{black}

\subsection{Wind Data Sources}

Unlike FRP and NDVI, surface wind speed and direction are not directly measurable through optical satellite imagery. Although atmospheric motion vectors can be derived by tracking cloud features in satellite image sequences, such as those using infrared or water vapor channels, these are indirect estimations typically applicable to higher atmospheric levels and often lack the direct surface-level precision necessary for fire modeling. Consequently, precise local wind information pertinent to wildfire propagation must be sourced from other dedicated meteorological resources, \cite{wang2014improved, bedka2005application}. 

To estimate local wind fields relevant to wildfire propagation, two primary data sources are generally utilized. First, Numerical Weather Prediction (NWP) models provide three-dimensional, gridded wind speed and direction fields at various altitudes and time intervals. Well-established global products, such as ERA5 (a reanalysis product from the European Centre for Medium-Range Weather Forecasts) and the Global Forecast System (GFS, providing operational forecasts from the National Oceanic and Atmospheric Administration), offer hourly or three-hourly wind estimates that can be spatially extracted for any point of interest. For fires occurring over land, these NWP datasets are widely regarded as the most reliable source for consistent and spatially comprehensive wind information that drives anisotropic fire spread models \cite{poblet2020era5, xu2024attention}. 

Alternatively, satellite scatterometers, such as ASCAT (on the MetOp series of polar-orbiting satellites), provide direct, all-weather observations of near-surface ocean wind speed and direction by measuring radar backscatter from the sea surface. However, these instruments are primarily designed to monitor oceanic wind fields and are therefore most applicable to fires near coastlines or on islands where their oceanic coverage extends over land or directly impacts coastal fire behavior. In practice, scatterometer data can complement NWP products by validating or refining near-surface wind patterns specifically over water bodies or coastal regions \cite{ricciardulli2021intercalibration, mandel2011wrf}. 

Combining high-resolution NWP wind estimates with scatterometer observations, when available, can enhance the accuracy of wind inputs used in fire spread modeling, thereby supporting more realistic simulations of directional fire behavior. Table \ref{winddataset} presents some of the common sources for wind datasets used in wildfire modeling \cite{poblet2020era5, xu2024attention, ricciardulli2021intercalibration, mandel2011wrf, knmi2018ascat}.

\begin{center}
\small
\begin{longtable}{@{}llll@{}}
\caption{Common wind data sources}\label{winddataset}\\
\toprule
\textbf{Dataset/Model} & \textbf{Type} & \textbf{Frequency} & \textbf{Wind Resolution} \\
\midrule
\endhead
\textbf{ECMWF ERA5} & Reanalysis (NWP) & Hourly & $\sim$31 km (global) \\
\textbf{GFS (NOAA)} & Forecast (NWP) & 3-hourly & $\sim$25 km (global) \\
\textbf{WRF} & Regional Model & User-defined & 1--10 km (typical) \\
\textbf{ASCAT (MetOp-A/B/C)} & Scatterometer & 2--4 passes/day & 12.5 km -- 25 km (ocean) \\
\textbf{SeaWinds (QuikSCAT)} & Scatterometer (historic) & 1--2 passes/day & 25 km (ocean) \\
\bottomrule
\end{longtable}
\end{center}

In the next section, we recall essential concepts, relationships, formulas, and tools from mathematics and physics to ensure that the paper is sufficiently self-contained.

\section{Estimating Rate of Spread}
\label{sec:ros}

Rate of spread is one of the most fundamental parameters for quantifying wildfire dynamics, as it determines how quickly the fire front advances across the landscape. Because, rate of spread depends on the interaction of multiple drivers—fuel moisture, wind, topography, and fire intensity—its estimation requires the integration of multisource satellite and meteorological datasets within a consistent modeling framework.

To estimate the rate of spread, we employ thermal observations and atmospheric variables to characterize the spatiotemporal evolution of the active fire front. Specifically, we use FRP data to quantify fire intensity and combine thermal imagery from a geostationary satellite with high temporal resolution (e.g., GOES) with fire perimeter information from polar-orbiting satellites with higher spatial resolution (e.g., VIIRS or MODIS). The geostationary platform provides continuous monitoring of thermal anomalies, capturing the rapid progression of active fires, while polar-orbiting instruments deliver detailed spatial information on fire boundaries and surrounding vegetation.

Atmospheric datasets are incorporated to derive the effective wind direction, denoted as $\hat{\theta}$, and wind speed, $U$. Combined with vegetation and fuel information, these inputs enable a spatially explicit estimation of anisotropic fire spread parameters, which are then used in Eq.~\eqref{ind} to obtain the directional rate of spread function. This approach provides the basis for modeling both head and backfire propagation under variable environmental conditions.

\subsection{Directional Rate of Spread}
\label{subsec:directional_ros}

The fire front represents the interface between burned and unburned fuel, delineating the active boundary where combustion progresses into new areas. Its geometry and propagation are influenced by factors such as wind and fuel characteristics~\cite{dehkordi2024approximated}. Although determining the exact fire front under real conditions is highly complex due to these interacting factors, approximate analytical descriptions can be obtained under simplified, quasi-stationary scenarios.

In particular, when a fire originates from a small ignition area under approximately constant environmental conditions (wind, humidity, topography, and fuel type), the fire front after a given time interval (e.g., 10–60 minutes) can be approximated by a closed curve, often represented as a \textit{deformed ellipse}~\cite{dehkordi2024approximated}. Let $R_H(x, y, t)$ and $R_B(x, y, t)$ denote, respectively, the head fire (fastest) and backfire (slowest) rates of spread at a given location $(x, y)$ and time $t$, and $U(x, y, t)$ the local wind speed at time $t$ and location $(x,y)$, where $x$ and $y$ are the latitude and longitude of the point. Then, the parameters $a(x, y, t)$, $b(x, y, t)$, and $c(x, y, t)$ are defined as:
\begin{equation} 
a = \frac{1 + 0.25U}{2(R_H + R_B)}, 
\quad 
b = \frac{1}{2}(R_H + R_B), 
\quad 
c = \frac{1}{2}(R_H - R_B).
\end{equation}
Let $\hat{\theta}(x, y, t)$ represent the wind direction and $\tilde{\theta}(x, y, t)$ the direction of maximum spread at time $t$ and point $(x,y)$. The fire front after a specified duration can then be expressed as \cite{dehkordi2024approximated}:
\begin{equation}
r(\theta, x, y, t) = 
\frac{ab}{
\sqrt{
a^2\cos^2(\theta - \hat{\theta} - \tilde{\theta}) + 
b^2\sin^2(\theta - \hat{\theta} - \tilde{\theta})
}} (x, y, t)
+ c\cos(\theta - \hat{\theta} - \tilde{\theta})(x, y, t).
\label{ind}
\end{equation}

Equation~\eqref{ind} generalizes the classical elliptical model used in the Rothermel formulation~\cite{rothermel1972mathematical, anderson1982modelling}, allowing for anisotropy induced by vegetation structure, atmospheric forcing, and terrain slope. This parametric representation forms the geometric foundation for our estimation of directional fire propagation, distinguishing between the head fire and backfire components.

In  Eq.~\eqref{ind}, the primary propagation direction, denoted by $\tilde{\theta}$, is estimated when observational data allows. However, if a direct and reliable estimation of the primary propagation direction is not available, the term $\tilde{\theta}$ is effectively canceled or set to zero in the calculation.

\subsection{Huygens' Principle}

Huygens' principle is a classical concept in wave theory that describes how wavefronts propagate through space. It states that every point on a given wavefront acts as a source of secondary wavelets, which expand in all directions at a speed determined by the local propagation rate. The new wavefront at a later time is then defined as the {envelope} of all these secondary wavelets~\cite{anderson1998modeling}. This geometric interpretation provides a natural framework for modeling the advance of a fire front, which behaves analogously to a propagating wave.

Wildfire propagation is inherently anisotropic due to the influence of wind, slope, and fuel heterogeneity. To account for these effects, we replace the circular wavelets in Huygens’ construction with the anisotropic spread shape described by $r(\theta, x,y)$, Eq.~\eqref{ind}~\cite{dehkordi2022applications}. 
Formally, if \( \Gamma(t) \) denotes the fire front at time \( t \), the fire perimeter at a later time \( t + \Delta t \) can be described as
\[
\Gamma(t + \Delta t) = \mathrm{Envelope}\!\left(\bigcup_{(x,y) \in F(t)} 
\left\{\, r(\theta, x,y,t)\, \Delta t \,\right\}\right),
\]
where  \(r(\theta, x,y,t)) \) is the \textit{directional rate of spread} at location \( (x,y)\), direction \( \theta \) and time $t$.

In the simplest isotropic case, \( r(\theta, x,y,t) \) is constant and the expanding wavelets are circular, corresponding to a uniform spread in all directions. However,
In this generalized formulation, at each time $t$, \( r(\theta, x,y,t) \) varies with both position and direction, producing a more realistic representation of the evolving fire front.

In the following section, we describe how the directional rate of spread \(r(\theta, x,y,t) \) is derived from satellite-based thermal and then refined and predicted using wind, and LFMC.


\subsection{Estimation Rate of Spread from Thermal Datasets}

Thermal satellite imagery provides sequential observations of active fire fronts, enabling the estimation of the fire’s directional expansion over time. By tracking the evolution of the front between consecutive detections, we derive empirical estimates of the head-fire ($R_H$) and backfire ($R_B$) rates of spread.

Let $\Gamma(t)$ denote the fire front at time $t$, defined from thermal imagery as the envelope enclosing all detected hotspots. To quantify fire progression, we examine consecutive thermal observations of the same region. When active fire pixels are detected at time $t_0$ and additional hotspots appear nearby at $t_0 + \Delta t$, the fire front is inferred to have advanced from the initial to the adjacent area within the interval $\Delta t$. By mapping all pixels that transition from unburned to burned throughout the time series, the evolving fire perimeter $\Gamma(t)$ can be reconstructed. This dynamic boundary describes both the geometry and the velocity of fire spread, forming the basis for estimating directional rates of spread from satellite data.

By linking active fire detections across consecutive timestamps, the fire front displacement can be computed. The empirical rate of spread in direction $\theta$ is then given by
\begin{equation}
R^T(\theta) = \frac{d(\theta)}{\Delta t},
\end{equation}
where $d(\theta)$ represents the maximum displacement of the fire front along direction $\theta$ between two consecutive detections separated by $\Delta t$. The directional components of this rate correspond to $R^T(\hat{\theta}) = R_H^T$ for the downwind (head-fire) direction and $R^T(\hat{\theta}+\pi) = R_B^T$ for the upwind (backfire) direction.

\begin{figure}[H]
    \centering
       \begin{subfigure}[b]{0.45\textwidth}
        \centering
        \includegraphics[width=.5\linewidth]{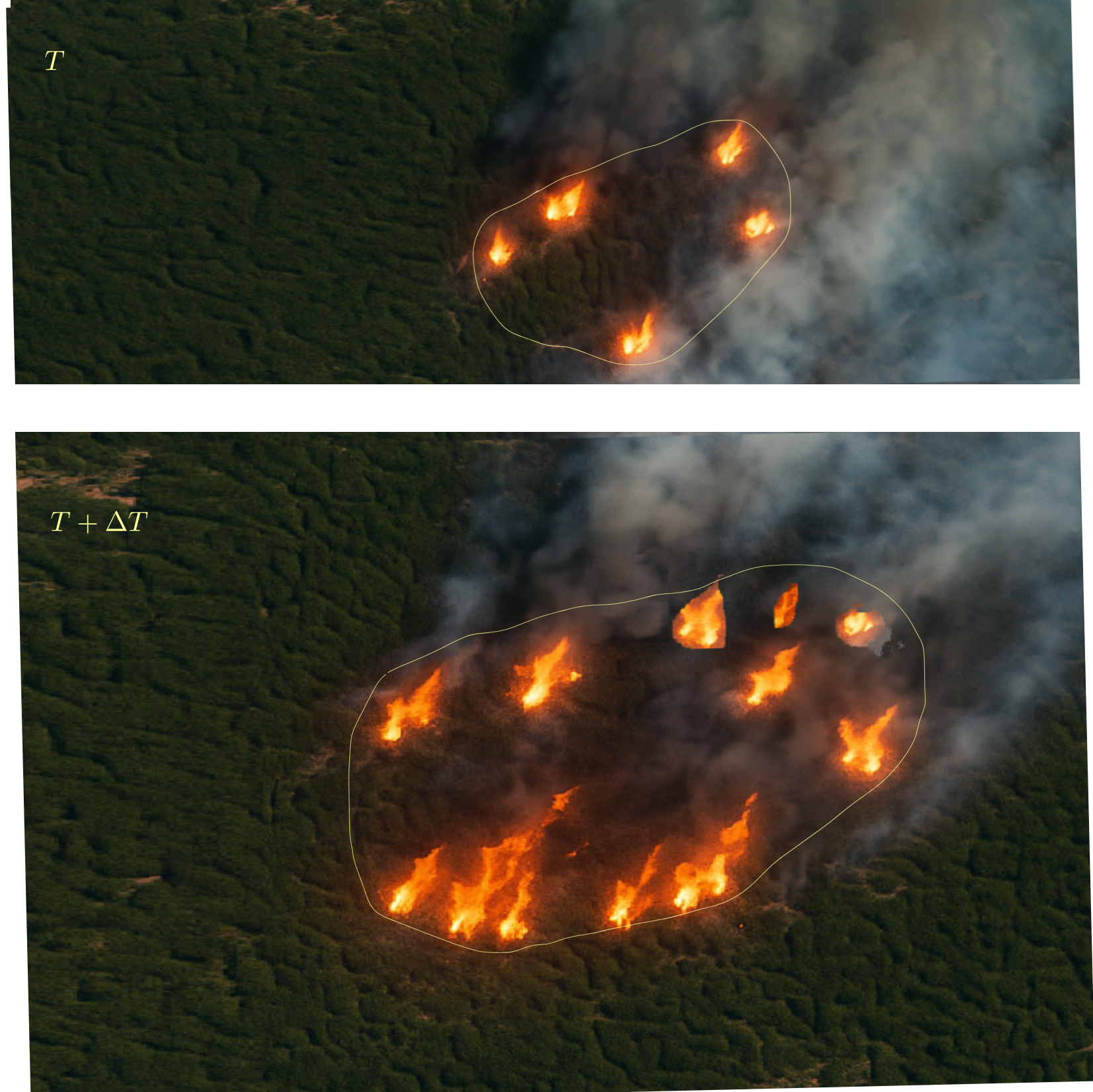}
        \subcaption{AI-generated image showing the fire front at time $T$ and after an interval $\Delta t$.}
        \label{fire1}
    \end{subfigure}
    \hfill
    \begin{subfigure}[b]{0.5\textwidth}
        \centering
        \includegraphics[width=.5\linewidth]{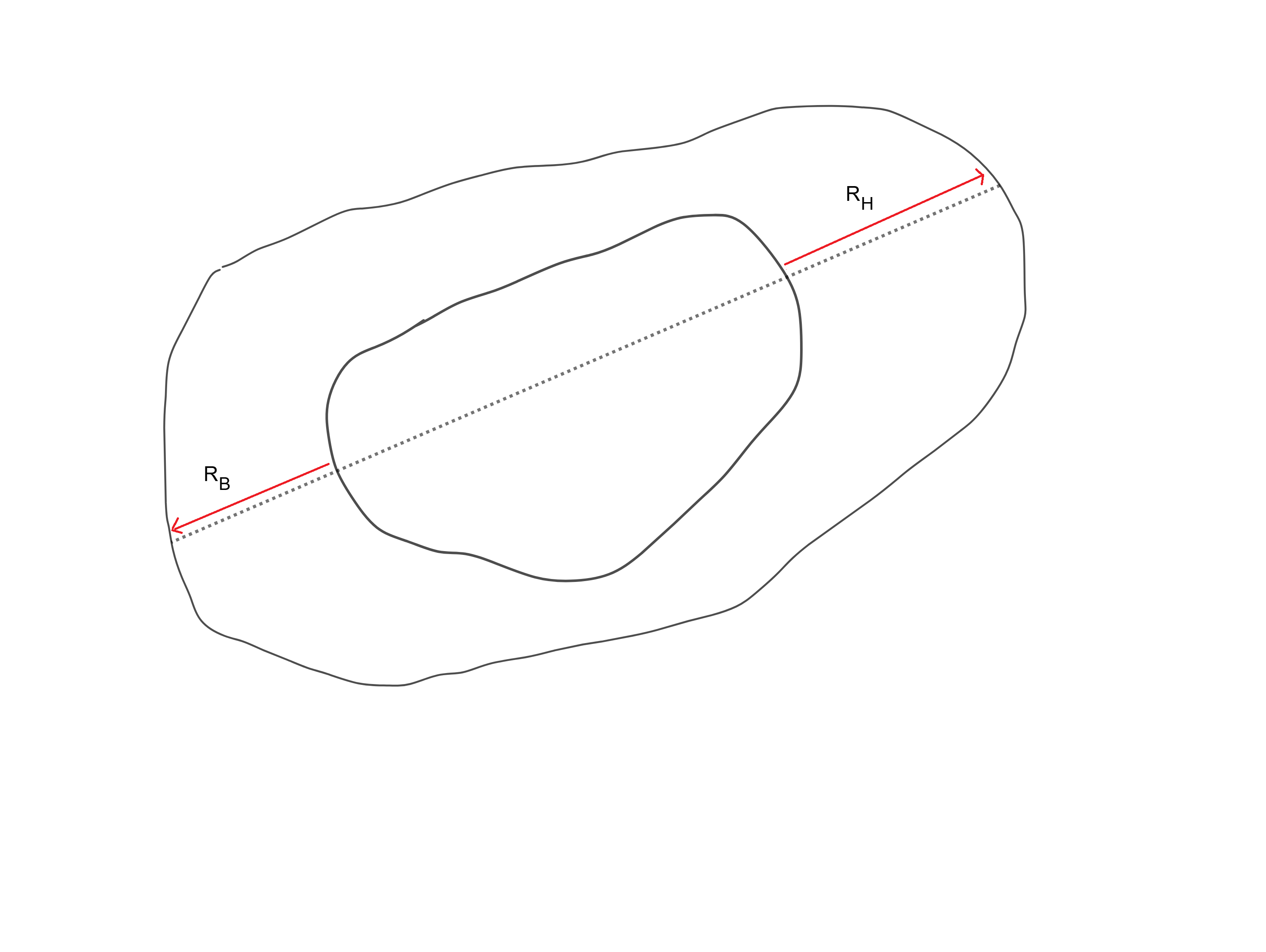}
\subcaption{Fire fronts depicted in the image (AI-generated image).}
        \label{fire2}
    \end{subfigure}
    \centering
    \begin{subfigure}[b]{0.55\textwidth}
        \centering
        \includegraphics[width=.55\linewidth]{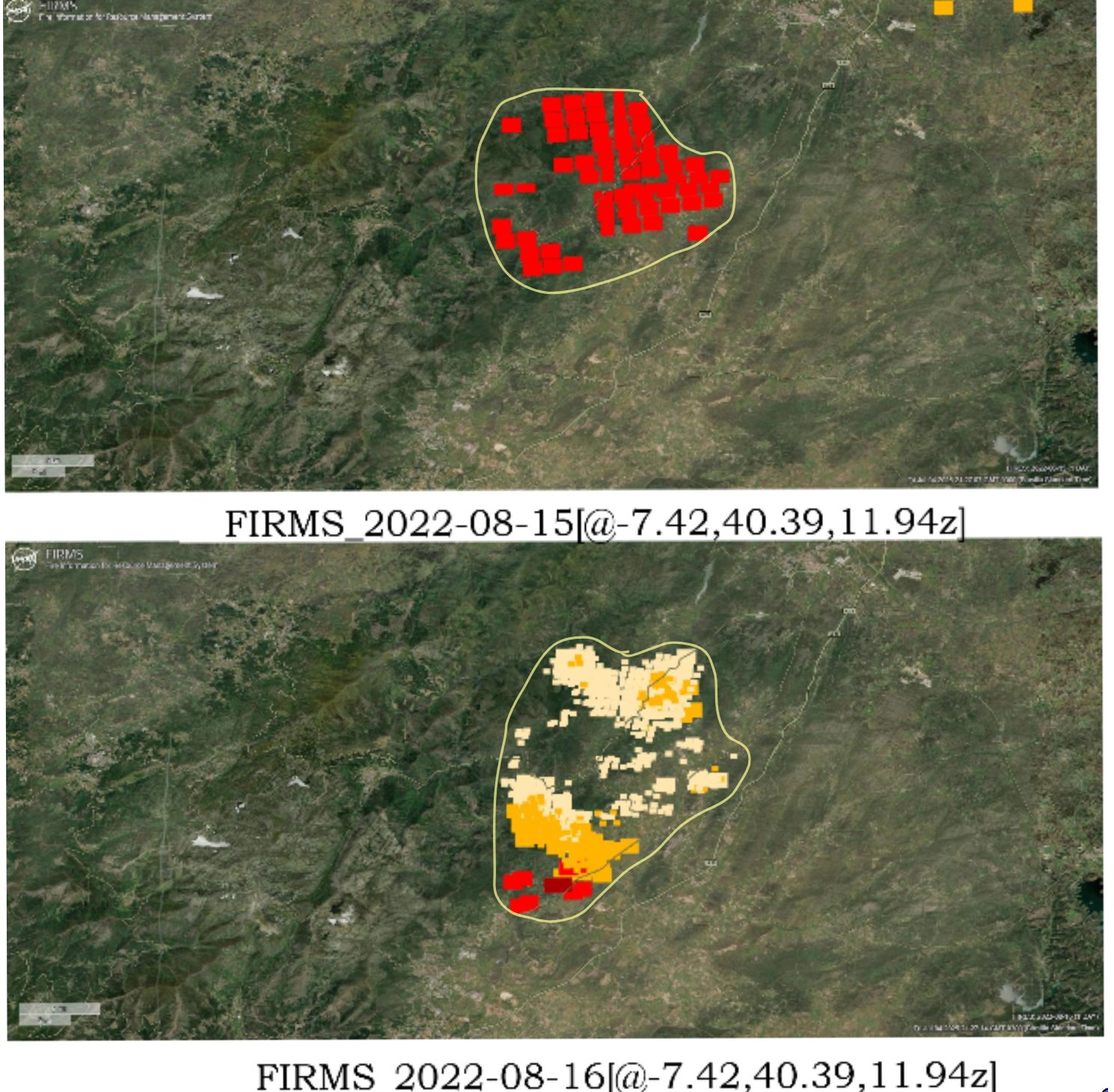}
\subcaption{Wildfire images in Portugal (NASA, 2022)}
        \label{fi1}
    \end{subfigure}
    \hfill
    \begin{subfigure}[b]{0.4\textwidth}
        \centering
        \includegraphics[width=.4\linewidth]{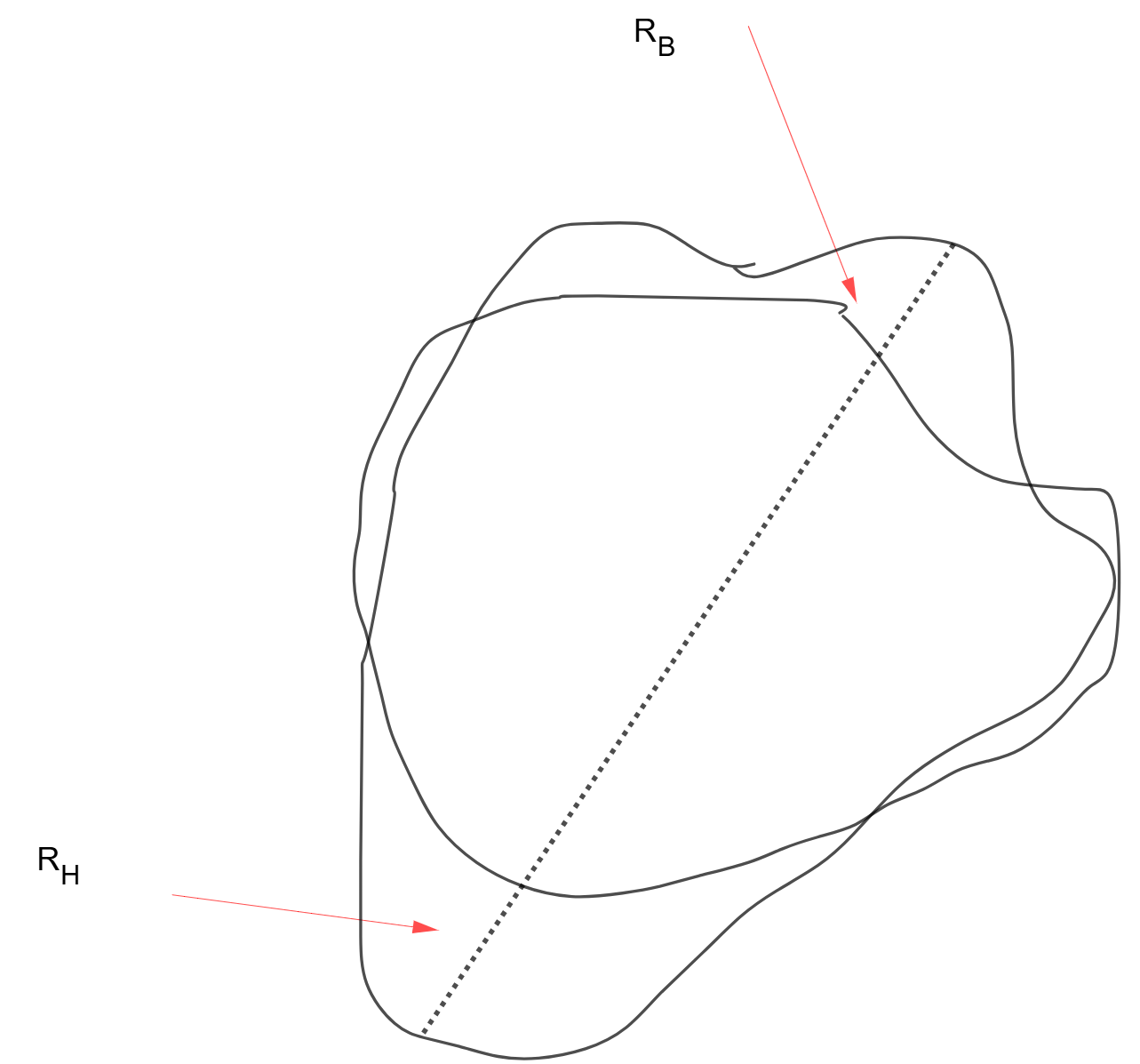}
\subcaption{Fire fronts depicted in the image (NASA images).}
        \label{fi2}
    \end{subfigure}
    \caption{Illustration of the procedure for extracting fire front positions and corresponding rates of spread from satellite imagery. }
    \label{f2}
\end{figure}

Figures~\ref{f2} illustrate two examples demonstrating the evolution of fire fronts and the estimation of their rates of spread. Figure~\ref{fire1} presents an AI-generated conceptual illustration of an idealized fire front expansion over time, highlighting the identification of head-fire and backfire rates. Figure~\ref{fi2} shows real satellite observations of wildfires that occurred in Portugal in August 2022. The thermal anomalies and burned areas captured on consecutive days clearly depict the progression of the fire and serve as input for estimating directional rates of spread.


In the next section, we describe how the thermal rates of spread, $R_H^T$ and $R_B^T$, are refined using satellite data and numerical methods to predict the head-fire and backfire rates of spread, $R_H$ and $R_B$, required in Eq.~\eqref{ind}.

\subsection{Numerical Method-Based Refinement of Thermal Rates of Spread }

The rate of fire spread depends not only on meteorological conditions but also on the spatial variability of fuel characteristics, which are closely linked to vegetation type and moisture content. Although thermal satellite imagery provides empirical estimates of directional rates of spread, these estimates do not explicitly account for local variations in fuel availability or environmental conditions. To overcome this limitation, we introduce correction functions that adjust the thermal-derived rates of spread, $R_H^T$ and $R_B^T$, to predict the head-fire and backfire rates of spread, $R_H$ and $R_B$, respectively. This correction incorporates satellite-based wind and vegetation data, thereby integrating both atmospheric and fuel influences into the model.

The refinement process begins with the estimation of $R_H^T$ and $R_B^T$ from two consecutive thermal satellite images. Each point along the fire front is then associated with the corresponding LFMC  and wind conditions obtained from satellite datasets. Using these variables, we predict the corrected rates of spread, $R_H$ and $R_B$, through numerical fitting techniques. The resulting values are substituted into the generalized elliptical growth model (Eq.~\ref{ind}), and the Huygens principle is applied to compute the subsequent fire front. 

This newly generated front provides the basis for the next iteration: LFMC and wind data are resampled around the updated perimeter, and refined values of $R_H$ and $R_B$ are recalculated. By iteratively repeating this procedure, a sequence of successive fire fronts is reconstructed, capturing the dynamic evolution of wildfire propagation under spatially varying vegetation and environmental conditions.

\subsubsection{Exponential relation between rate of spread, wind, and fuel moisture}

We adopt the widely used multiplicative form with an exponential damping for moisture \cite{rossa2017effect, cheney1998prediction, rossa2016laboratory, rossa2018effect}:
\begin{equation}\label{ros_exp}
    R(x,y,t) \;=\; \mathbf{A}\,U^{\alpha}(x,y,t)\,\exp\big(-\beta\,M(x,y,t)\big),
\end{equation}
where
\begin{itemize}
    \item $R(x,y,t)$ is the head fire (or backfire) rate of spread at location $(x,y)$ (m/min);
    \item $U(x,y,t)$ is the relevant wind speed at time $t$ and location $(x,y)$ (m/s);
    \item $M(x,y,t)=LFMC(x,y,t)$ is the fuel moisture content at time $t$, and $(x,y)$ (in \%);
    \item $\mathbf{A}>0$, $\alpha>0$, and $\beta>0$ are empirical parameters to be calibrated for the study region. We emphasize that $\beta > 0$ leads to the expected reduction of $R$ with increasing moisture content $M$.

\end{itemize}

Let $R_H^T(x,y, t)$ ($R_B^T(x,y, t)$) denotes the head fire (backfire) rate of spread obtained from the thermal satellite observations (or a reference model) at the detection/measurement time and location observed under local conditions $(U_{{sat}}(x,y, t),M_{{sat}}(x,y, t))$, and we wish to predict the head fire (backfire), $R_H (R_B)$, under regional conditions $(U_{{reg}}(x,y, t),M_{{reg}}(x,y, t))$. Then, by EQ.~\eqref{ros_exp}, the correction between the thermal head fire rate of spread, $R^T_H$, and predicted head fire rate of spread, $R_H$, is
\begin{equation}\label{correction_head}
R_H(x,y, t) \;=\; R_H^T(x,y, t)\;
\left(\dfrac{U_{{reg}}(x,y, t)}{U_{{sat}}(x,y, t)}\right)^{\!\alpha_H}
\exp\!\Big(-\beta_H\,(\,M_{{reg}}(x,y, t)-M_{{sat}}(x,y, t)\,)\Big).
\end{equation}
We have a similar equation for the thermal backfire rate of spread, $R^T_B$, and predicted back fire rate of spread, $R_B$. If the local moisture exceeds a combustion sustainment threshold $M_{\min}$ (or if fuels are effectively non-combustible), set the corrected rates of spread to zero:
\[
R_{H,B}(x,y, t)=0 \quad\text{if } M_{{reg}}(x,y, t)\ge M_{\min}.
\]


EQ.~\eqref{correction_head} multiplies the satellite-derived rate of spread by a local correction factor that depends on the ratio of wind speeds and on the difference in moisture between the target and the satellite-measured conditions. The important point is that the coefficients $\alpha$ and $\beta$ are different for backfire and head fires and the reason is that the head fires is more sensitive to the wind and the backfire is more sensitive to the moisture because it is less sensitive to the wind. To find the values of $\alpha$ and $\beta$,  we can choose of these strategies: 
\begin{enumerate}
\item Pick plausible values (or ranges) for $\alpha_H$ and $\beta_H$ from the literature, compute $R_{H}$ for each combination, and report the resulting range. Several studies have employed laboratory or field experiments to determine constants in power-law or exponential relationships linking the rate of spread to fuel moisture content and wind speed. However, to the best of our knowledge, none explicitly report the precise ranges of $\alpha$ or $\beta$. Based on the available studies, the following intervals can be inferred. For instance, in Figure~1 of \cite{xavier1998forest}, the author investigated \textit{Pinus pinaster} needle fuels to establish the relationship between the rate of spread and fuel moisture content, yielding $\beta$ values within $[0.05,\,0.07]$. Typical choices for $\alpha$ are $[1.0,\,1.5]$ (grass), $[0.8,\,1.2]$ (shrub), and $[0.3,\,0.8]$ (forest litter), as reported in Tables~2,~3, and~28 of \cite{andrews2018rothermel}. For $\beta$, some of the representative intervals are $[0.05,\,0.1]$ for dry or cured fuels, $[0.1,\,0.15]$ for moderately moist fuels, and $[0.15,\,0.25]$ for live or green fuels \cite{xavier1998forest, rossa2018empirical}. This approach is particularly suitable for regional-scale fire modeling, where vegetation types are relatively homogeneous and empirical parameter estimates from previous local studies are available.

\item Collect several matched pairs $(R, U, M)$ and estimate $\alpha$ and $\beta$ from data. Considering the model
\[
R = \mathbf{A}\, U^{\alpha}\, e^{-\beta M},
\]
taking logarithms yields the linear form
\[
\ln R = \ln \mathbf{A} + \alpha \ln U - \beta M,
\]
which allows direct calibration of $(\mathbf{A}, \alpha, \beta)$ through multiple linear regression using matched observations $\{(R_i, U_i, M_i)\}$ obtained from satellite thermal imagery and regional meteorological data. This data-driven method is more appropriate for global-scale analyses, where high biodiversity, varying vegetation structures, and extensive satellite coverage enable robust statistical estimation of model parameters across diverse ecosystems.
\end{enumerate}

\paragraph{Worked Numeric Example}

We now fix the thermal-derived rate of spread to $R_T=2.00\ \mathrm{m\,min^{-1}}$ (observed under
$U_{\mathrm{sat}}=4.0\ \mathrm{m\,s^{-1}}$ and $M_{\mathrm{sat}}=8\%$) and compute corrected head-fire
rates $R_H$ for several combinations of regional wind speeds $U_{\mathrm{reg}}$ and moisture contents
$M_{\mathrm{reg}}$ using Eq.~\eqref{correction_head}, with value obtained form the litarature as $\alpha=0.70$ and $\beta=0.039$.

Table~\ref{tab:RH_grid} reports $R_H$ for $U_{\mathrm{reg}}\in\{2,4,6,8\}\ \mathrm{m\,s^{-1}}$ and
$M_{\mathrm{reg}}\in\{6,8,10,12\}\%$.

\begin{table}[H]
\centering
\caption{Corrected head-fire rates $R_H$ (m\,min$^{-1}$) for fixed $R_T=2.00$, $U_{\mathrm{sat}}=4.0$, $M_{\mathrm{sat}}=8\%$, $\alpha=0.70$, $\beta=0.039$.}
\label{tab:RH_grid}
\begin{tabular}{cccccc}
\toprule
$U_{\mathrm{reg}}$ (m\,s$^{-1}$) & $M_{\mathrm{reg}}=6\%$ & $M_{\mathrm{reg}}=8\%$ & $M_{\mathrm{reg}}=10\%$ & $M_{\mathrm{reg}}=12\%$ \\
\midrule
2.0 & $1.331$ & $1.231$ & $1.139$ & $1.053$ \\
4.0 & $2.162$ & $2.000$ & $1.850$ & $1.711$ \\
6.0 & $2.872$ & $2.656$ & $2.457$ & $2.273$ \\
8.0 & $3.513$ & $3.249$ & $3.005$ & $2.780$ \\
\bottomrule
\end{tabular}
\end{table}

The table shows that:
  Increasing $U_{\mathrm{reg}}$ (row-wise) systematically increases $R_H$ via the factor $(U_{\mathrm{reg}}/U_{\mathrm{sat}})^{\alpha}$.
  Increasing $M_{\mathrm{reg}}$ (column-wise) reduces $R_H$ through the exponential damping $\exp[-\beta\,(M_{\mathrm{reg}}-M_{\mathrm{sat}})]$.
  When both wind and moisture rise, their effects combine nonlinearly; for example, going from $(U,M)=(4,8)$ to $(6,10)$ increases the wind factor but also applies a larger moisture damping, resulting in an intermediate net change.
Extending this computation to each point along a satellite-derived fire front yields spatially varying fields $R_H(x,y, t)$ and $R_B(x,y, t)$, which can then be incorporated into the generalized elliptical model Eq.~\eqref{ind} and the Huygens-based wavefront propagation scheme. In Appendix \ref{apen}, we recall the steps to provide the rate of spread and the waves of fire.

\subsection{Example: Eight Huygens Waves from Satellite-Derived Rates}

In this example we reconstruct eight successive fire fronts starting from an observed (satellite)
initial fire location and using the directional rate-of-spread data provided. The initial wave
$\Gamma_0$ is taken as the observed perimeter (outer curve in Fig.~\ref{fire2}). The thermal-derived
and corrected head-fire and back-fire rates for the seven subsequent propagation steps are

\[
\begin{aligned}
R_H &= [\,5,\; 5.9,\; 2,\; 7,\; 6,\; 6.3,\; 3\,],\\[4pt]
R_B &= [\,4,\; 4.2,\; 1.5,\; 6,\; 4,\; 5.2,\; 2\,],
\end{aligned}
\]
and the corresponding wind magnitudes and directions (degrees) are
\[
U=[3,\,5,\,2,\,4,\,3,\,5,\,7],\qquad
\widehat\theta=[20^\circ,\,15^\circ,\,30^\circ,\,45^\circ,\,30^\circ,\,60^\circ,\,30^\circ].
\]

\begin{figure}
    \centering
    \includegraphics[width=1\linewidth]{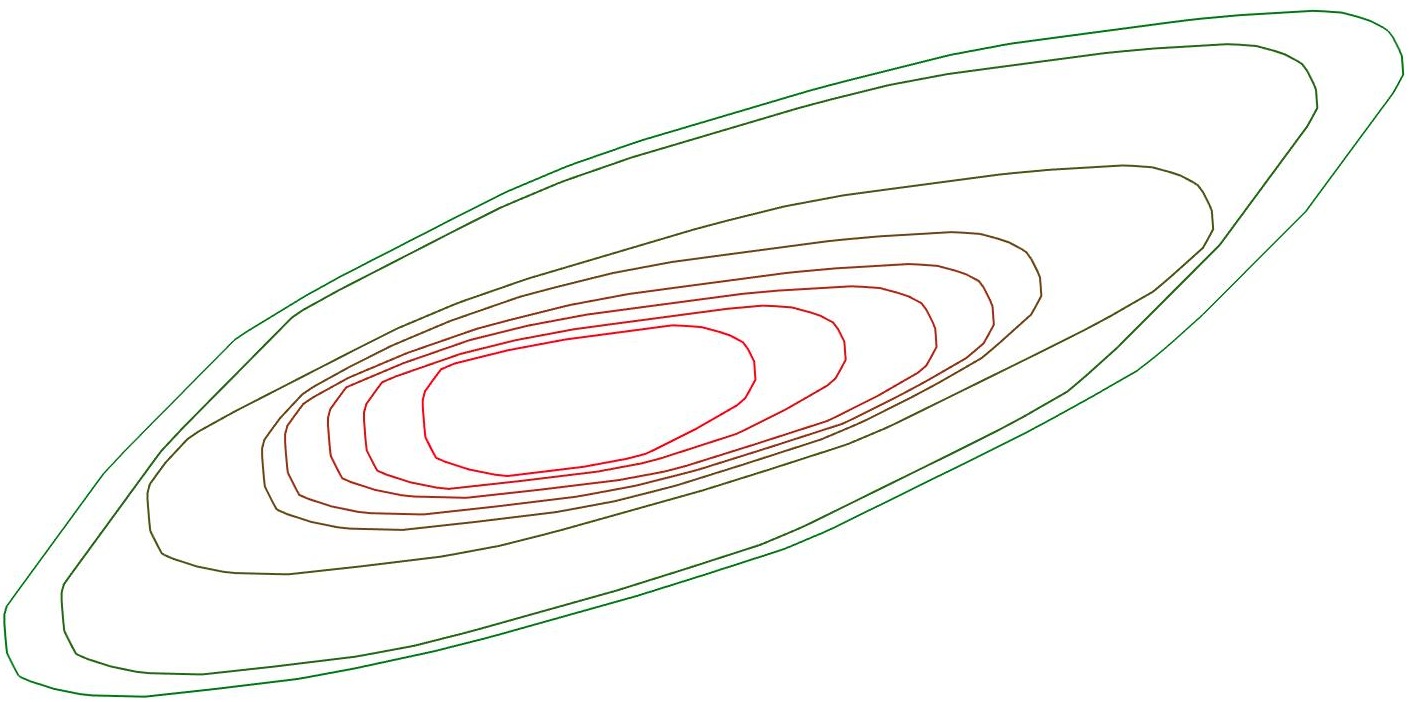}
    \caption{eight Huygens waves from satellite-derived rates}
    \label{8waves}
\end{figure}

The directional rate of spread, Eq.~\ref{ind}, is applied iteratively to estimate the fire fronts
for the following eight hours, assuming constant environmental conditions during each hourly step.
The result is shown in Fig.~\ref{8waves}, where each closed contour corresponds to a predicted
wavefront (or Huygens wave) of the fire perimeter.

The sequence of eight contours in Fig.~\ref{8waves} represents the temporal evolution of the fire
from its initially observed perimeter (innermost red contour) toward the most recent simulated
front (outermost green contour). Each wave $\Gamma_i$ corresponds to one propagation stage
computed from the respective $(R_H,R_B,U,\widehat{\theta})$ values.

At the first stage, the front expands mainly toward the northeast,
driven by a moderate wind ($U=3$~m\,s$^{-1}$, $\widehat{\theta}=20^\circ$) and
a slightly higher head-fire rate ($R_H=5$) than back-fire rate ($R_B=4$),
producing an elongated ellipse aligned with the wind.

During the second stage, the stronger wind ($U=5$) at a direction of $15^\circ$
intensifies the downwind spread, stretching the contour and increasing anisotropy.
The third stage corresponds to a calmer period ($U=2$) and smaller rates of spread
($R_H=2$, $R_B=1.5$), producing a more circular, compact wavefront.
At the fourth stage, the wind veers to $45^\circ$ and the head-fire rate rises
to $R_H=7$, generating a pronounced elongation in the northeast direction.

Subsequent waves exhibit the combined effect of varying wind and local fire intensity.
At the fifth and sixth stages ($\widehat{\theta}=30^\circ$ and $60^\circ$, respectively),
the fronts remain anisotropic but start to tilt, forming an oblique pattern consistent
with the shifting wind direction. The seventh and final stage corresponds to a stronger
wind ($U=7$, $\widehat{\theta}=30^\circ$), producing the outermost contour that advances
downwind with maximum elongation. The resulting configuration reproduces the characteristic
multi-lobed pattern of directional fire spread, where successive wavefronts are elongated
along the dominant wind axis and compressed upwind.

Overall, the figure illustrates how even moderate changes in the wind direction and speed
can produce the asymmetric nested structures of real wildfires. Each successive contour
represents the new envelope predicted by the Huygens propagation principle,
with the anisotropy governed by the input rates $(R_H,R_B)$ and the dynamic wind parameters
$(U,\widehat{\theta})$.

In addition to validating the model against observed fire perimeters, it is also important to understand how sensitive the propagation framework is to small changes in the environmental inputs that influence its directional behavior. To examine this aspect, we conducted a controlled numerical experiment in which we introduced slight variations in wind direction, wind velocity, or in the prescribed head-fire and back-fire rates of spread derived from the calibration procedure. Each modified configuration was then propagated using the same Huygens-based scheme described in this study, ensuring that any differences in the resulting wavefronts reflect solely the effects of these targeted parameter perturbations.

The comparison reveals that even modest adjustments in the driving wind field or directional rate of spread values can lead to changes in the predicted fire perimeter, particularly along the head-fire direction where the model demonstrates greater sensitivity. This behavior is consistent with the anisotropic structure of the rate of spread function and provides an additional qualitative assessment of the robustness of the proposed approach.

\begin{figure}[ht!]
    \centering
    \begin{subfigure}{0.48\textwidth}
        \centering
        \includegraphics[width=\linewidth]{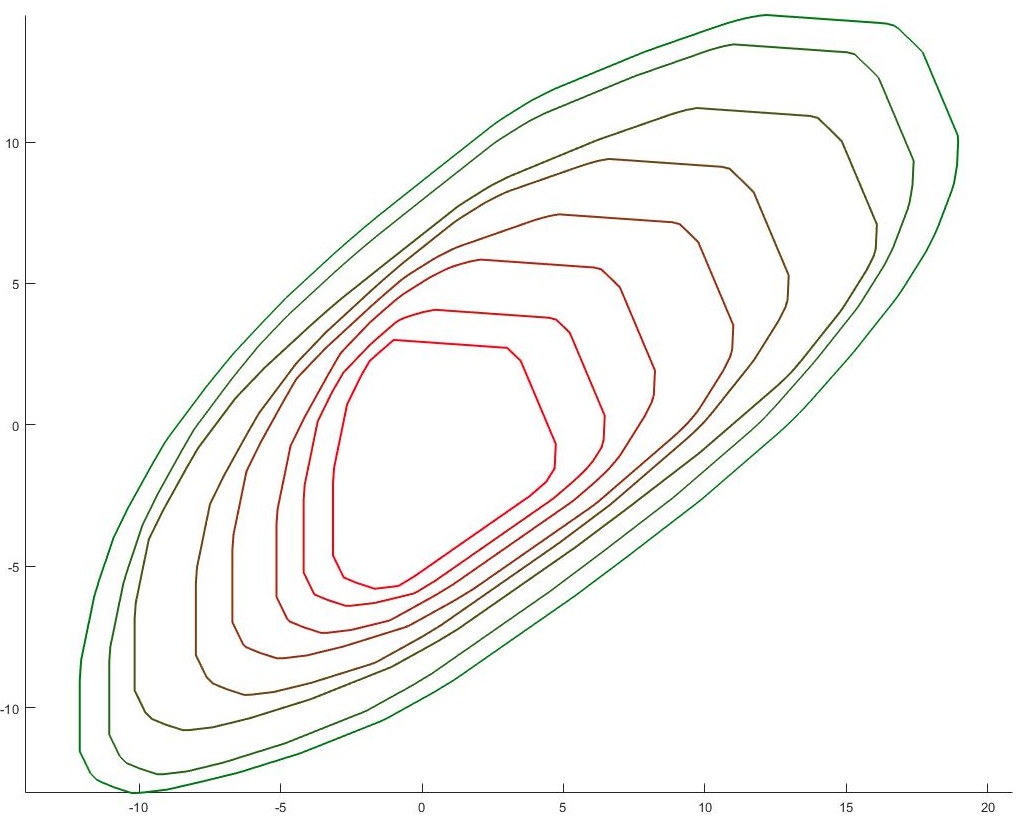}
        \caption{$R_H = [5, 5.9, 2, 7, 6, 6.3, 3]\\
R_B = [4, 4.2, 1.5, 6, 4, 5.2, 2]\\
U = [3, 5, 2, 4, 3, 5, 7]\\
\widehat\theta = [20^\circ, 15^\circ, 30^\circ, 45^\circ, 30^\circ, 60^\circ, 30^\circ]$
}
        \label{fig:waves2}
    \end{subfigure}
    \hfill
    \begin{subfigure}{0.48\textwidth}
        \centering
        \includegraphics[width=\linewidth]{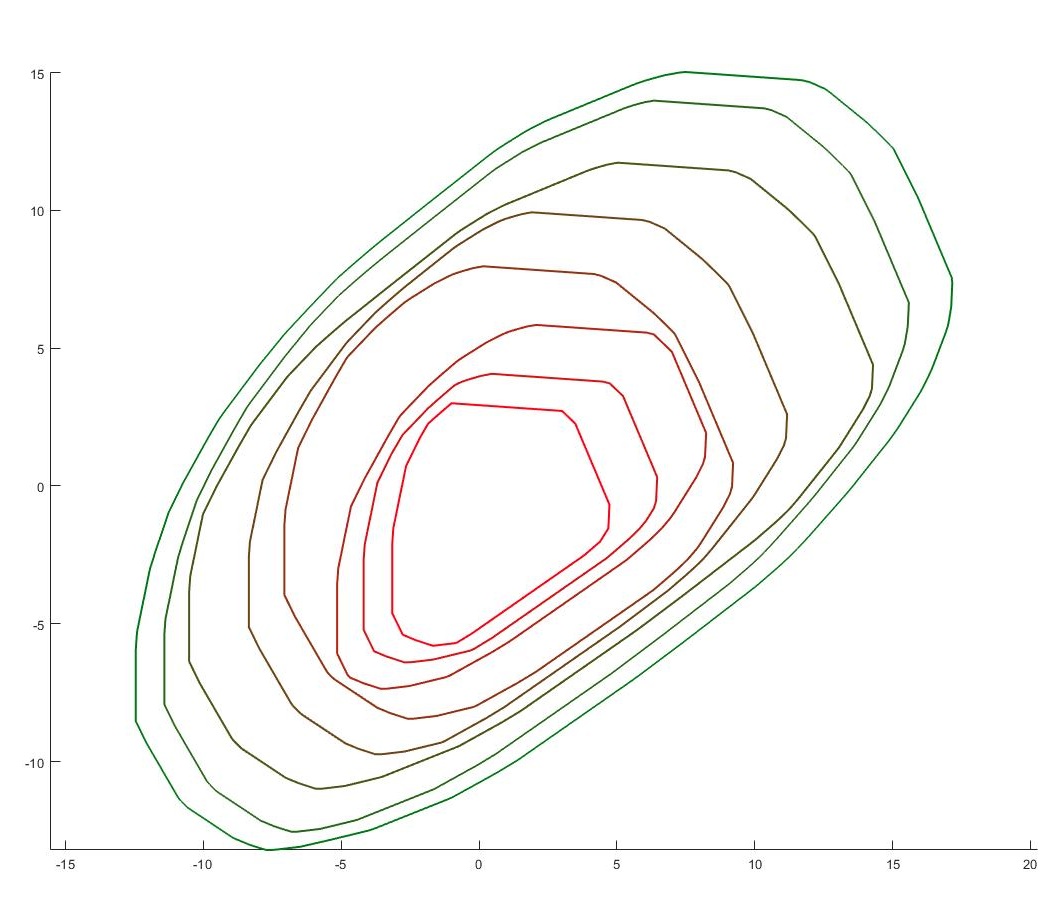}
        \caption{$R_H = [2, 2.5, 3.2, 3.5, 3.6, 2.6, 1.8]\\
R_B = [1.2, 1.4, 1.8, 2.6, 2.5, 1.8, 1.2]\\
U = [4, 5, 2, 4, 3, 5, 7]\\
\hat{\theta} = [30^\circ, 45^\circ, 30^\circ, 45^\circ, 30^\circ, 60^\circ, 30^\circ]$}
        \label{fig:waves3}
    \end{subfigure}

    \vspace{0.3cm}

    \begin{subfigure}{0.48\textwidth}
        \centering
        \includegraphics[width=\linewidth]{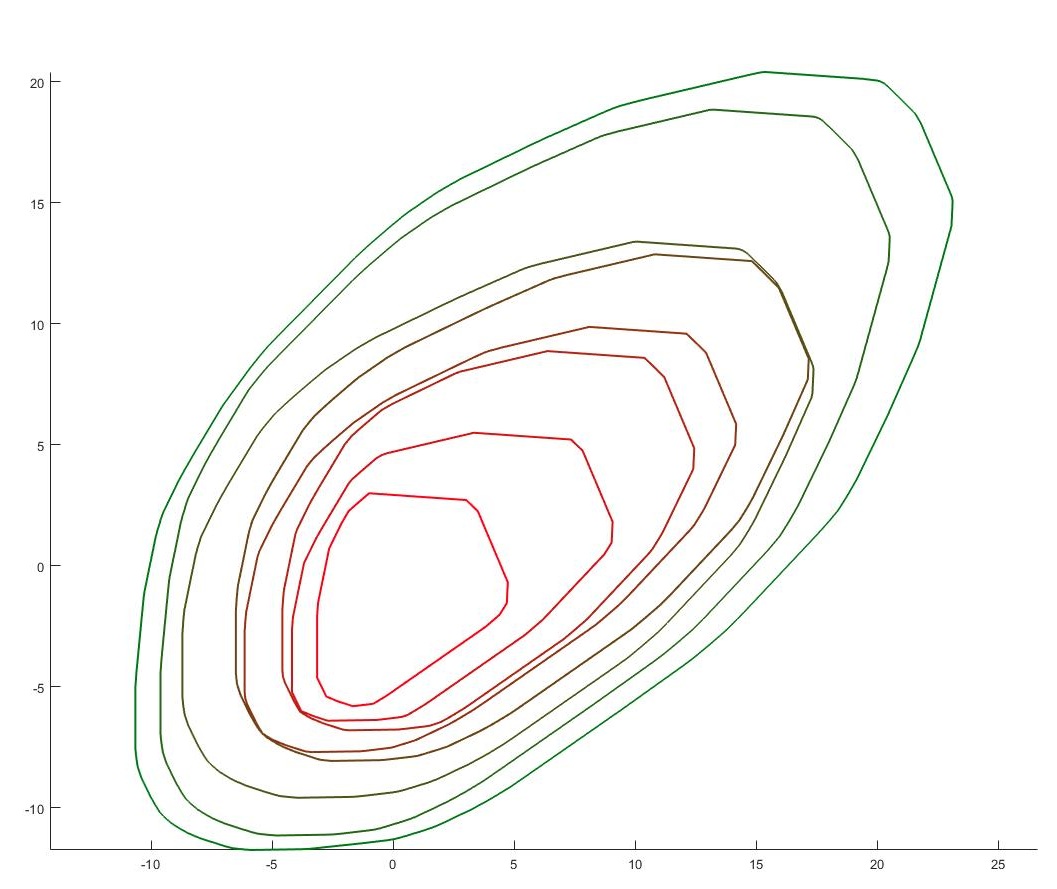}
        \caption{$R_H = [1.2, 1.4, 1.8, 2.6, 2.5, 1.8, 1.2]\\
R_B = [5, 5.9, 2, 7, 6s^{x+16}, 6.3, 3+e^{-3(y+14)}]\\
U = [4, 5, 2, 4, 3, 5, 7]\\
\hat{\theta} = [30^\circ, 45^\circ, 30^\circ, 45^\circ, 30^\circ, 60^\circ, 30^\circ]$}
        \label{fig:waves4}
    \end{subfigure}
    \hfill
    \begin{subfigure}{0.48\textwidth}
        \centering
        \includegraphics[width=\linewidth]{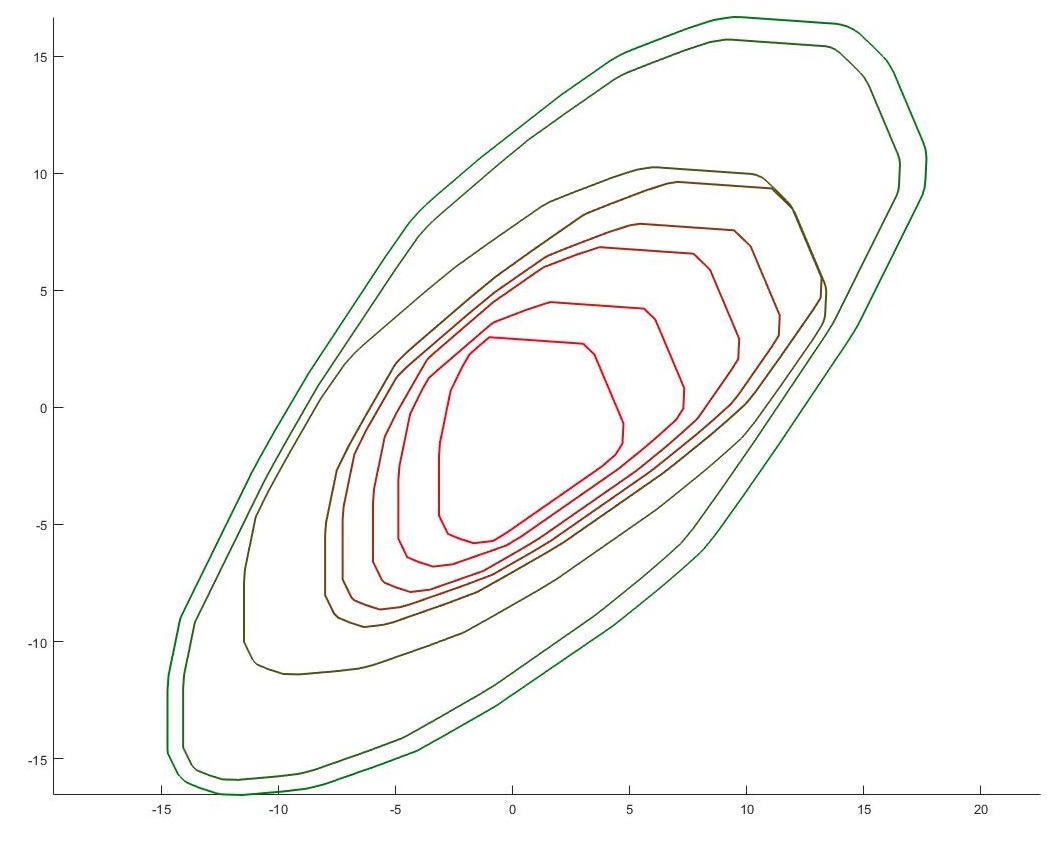}
        \caption{$R_H = [3, 3.9, 2, 6, 62^{x+16}, 6.3, 1.2+e^{-3(y+2)}]\\
R_B = [2, 2.2, 1.5, 4.6, 4+e^{-x^2}.2^{(y+16)/100}, 5.2, .7+e^{-3x^2}]\\
U = [4, 5, 2, 4, 3, 5, 7]\\
\hat{\theta} = [30^\circ, 45^\circ, 30^\circ, 45^\circ, 30^\circ, 60^\circ, 30^\circ]$}
        \label{fig:waves5}
    \end{subfigure}

    \caption{Comparison of wavefronts with small changes in conditions.}
    \label{port-waves}
\end{figure}

Fig.~\ref{port-waves} presents the four predicted wavefront sequences corresponding to the Portugal fire, specifically capturing the outermost contour of Fig.~\ref{fi2}. All wavefronts are plotted in a common coordinate system to enhance visual comparability, and the parameter values associated with each scenario are listed beneath the respective panels.

\subsection{Example: Modeling the Spread of the Eaton Fire using Generalized Elliptical Frames}

\begin{figure}[ht!]
    \centering
        \includegraphics[width=\linewidth]{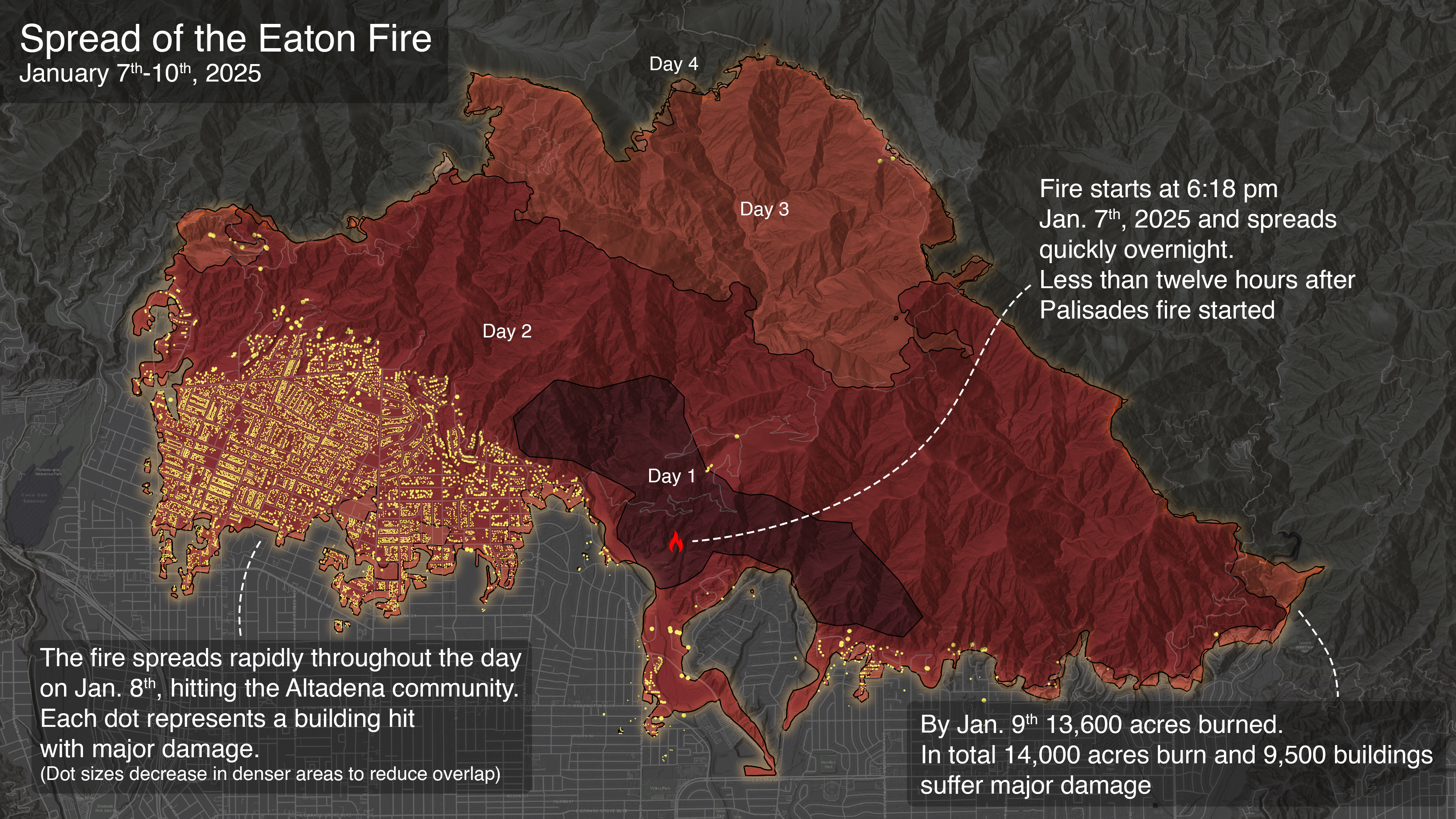}
\caption{Satellite imagery adapted from 
\href{https://svs.gsfc.nasa.gov/5568/\#media_group_378462}{NASA SVS}.}
        \label{eaton}
    \end{figure}

\begin{figure}[ht!]
    \centering
    \begin{subfigure}{0.4\textwidth}
        \centering
        \includegraphics[width=\linewidth]{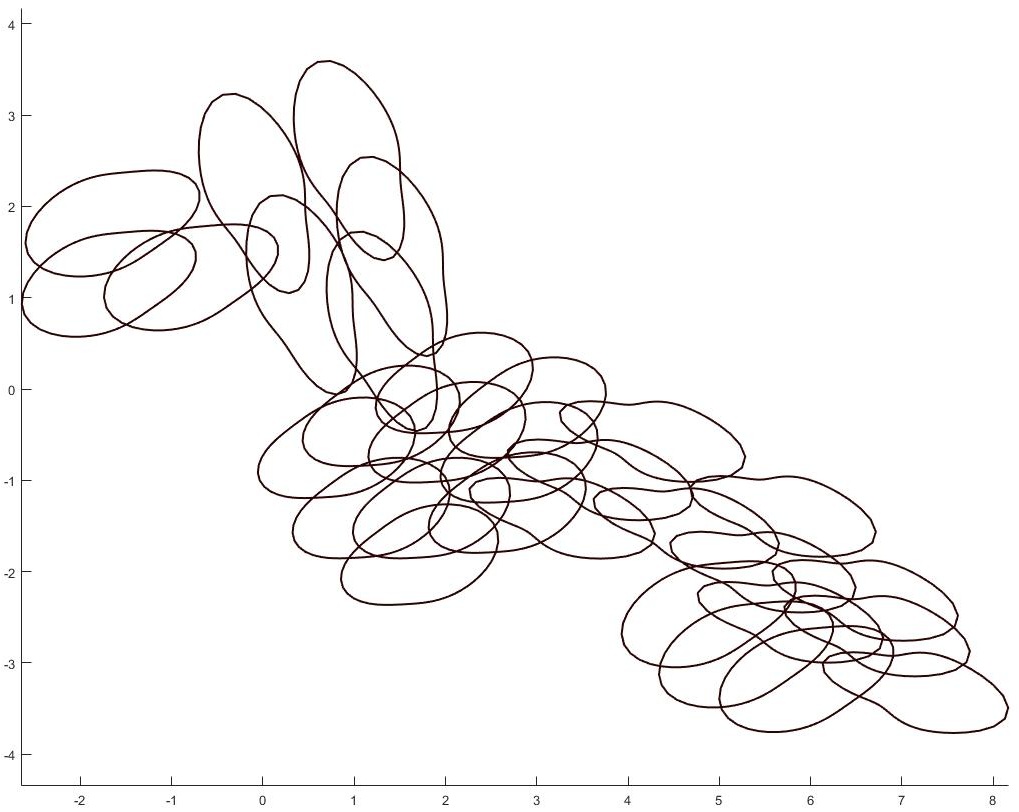}
        \caption{The propagation for the first day in which the wind changes several times. The Data are $R_H = [1 , 1.3, 1.2, 1.1]$, $R_B = [.8, 1, .9, .9]$, $U =  [3, 5, 2, 4]$, $\widehat\theta = [110^\circ, 200^\circ, 75^\circ , 290^\circ]$}
        \label{eaton1}
    \end{subfigure}
    \hfill
    \begin{subfigure}{0.48\textwidth}
        \centering
        \includegraphics[width=\linewidth]{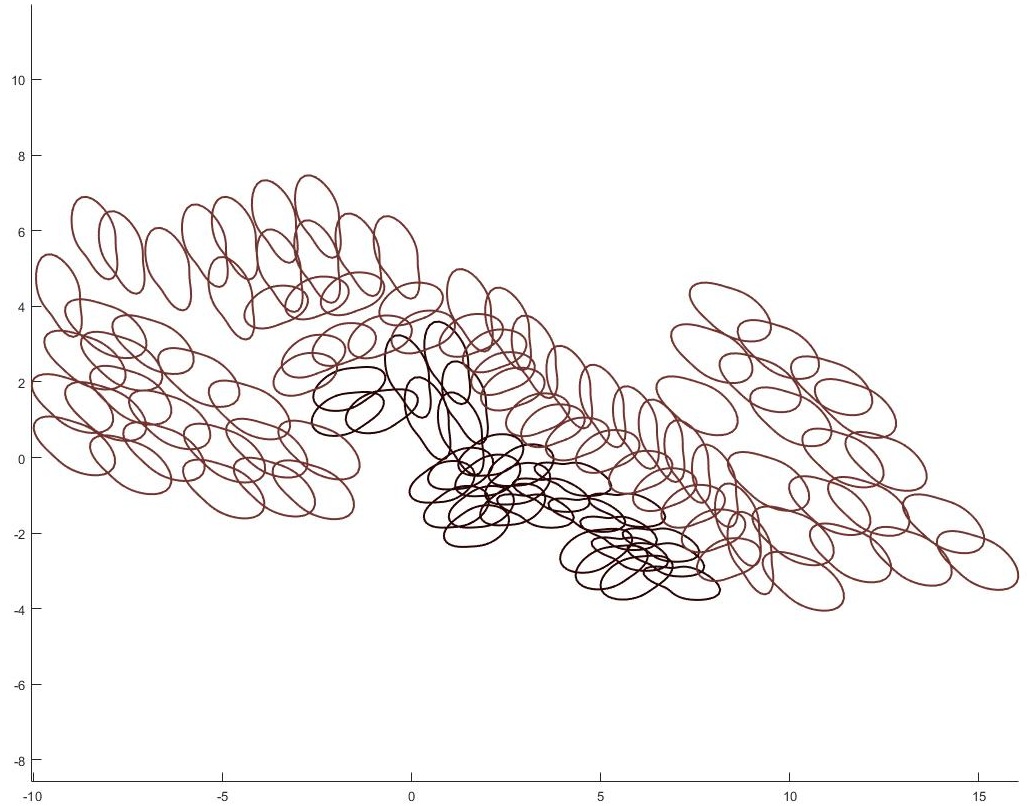}
        \caption{$R_H = [1 , 1.3, 1.2, 1.1, .6, 1.3, 1.3, 1.5 , 1.3]$, $R_B = [ .4, 1, 1.1 ]$, $U =  [ 3, 5, 7]$, $\widehat\theta = [ 200^\circ, 200^\circ, 60^\circ]$}
        \label{eaton2}
    \end{subfigure}

    \vspace{0.3cm}

    \begin{subfigure}{0.48\textwidth}
        \centering
        \includegraphics[width=\linewidth]{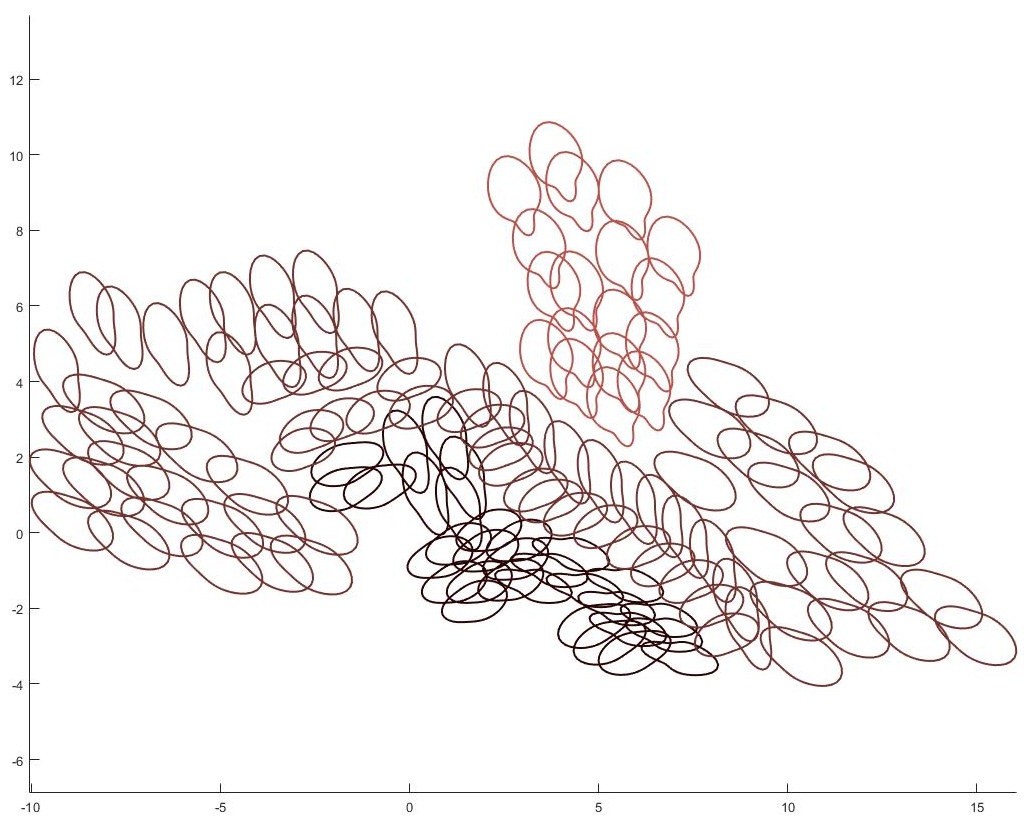}
        \caption{$R_H = [ 1.5 ]$, $R_B = [ .7 ]$, $U =  [5 ]$, $\widehat\theta = [ 200^\circ ]$}
        \label{eaton3}
    \end{subfigure}
    \hfill
    \begin{subfigure}{0.48\textwidth}
        \centering
        \includegraphics[width=\linewidth]{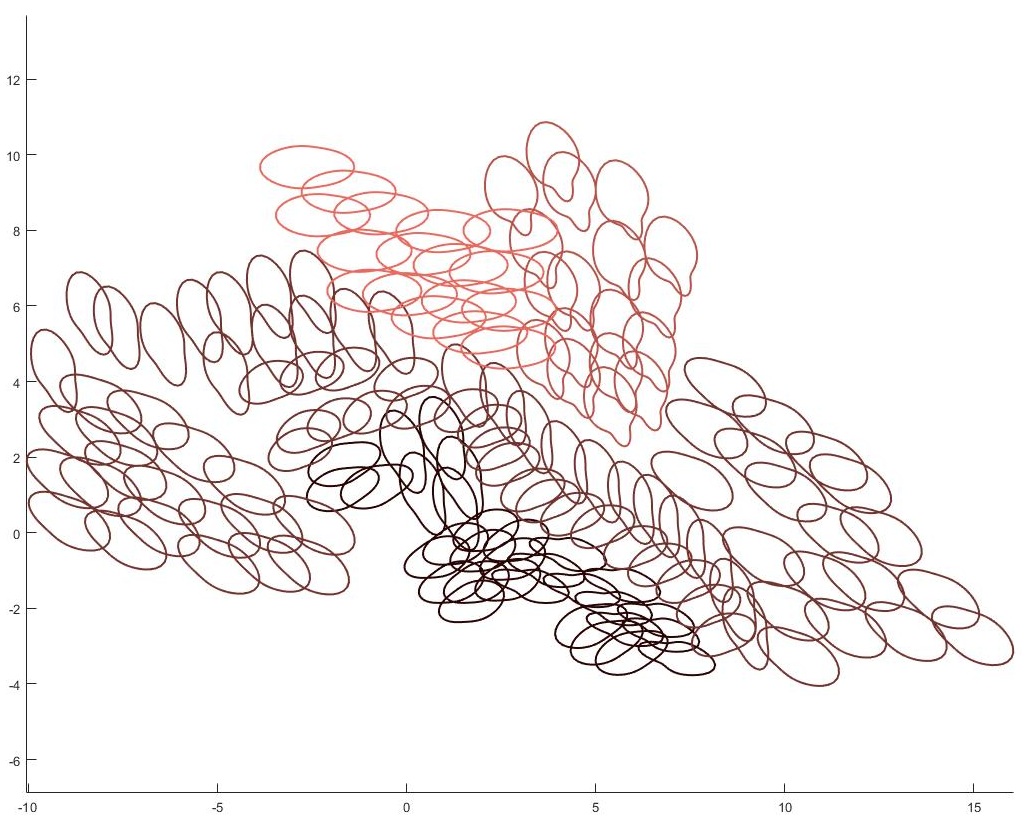}
        \caption{$R_H = [ 1.3]$, $R_B = [ 1.2]$, $U =  [ 7]$, $\widehat\theta = [270^\circ]$}
        \label{eaton4}
    \end{subfigure}

    \caption{}
    \label{eaton5}
\end{figure}

To demonstrate the practical relevance of the proposed generalized elliptical frame for wildfire propagation, we applied the model to the Eaton Fire (January 7–10, 2025). Fig.~\ref{eaton} shows the real progression of the fire over four days, alongside our model-generated propagation patterns derived from the generalized elliptical frames shown by Figs.\ref{eaton1}-\ref{eaton4}. Although detailed operational fire-spread data—such as measured wind velocity, wind direction, and local rates of spread—were not publicly available for this event, we constructed a plausible dataset that reflects the dominant environmental drivers: wind speed 
 relative to north, and head- and back-fire rates of spread. These values were selected within realistic ranges to reproduce the anisotropic growth expected under the reported meteorological conditions.

This example is not intended as a calibrated reconstruction of the Eaton Fire but rather as an illustrative case study demonstrating that the generalized elliptical frame framework can naturally capture complex, multi-day propagation patterns seen in real wildfires. Even with synthetic inputs, the model reproduces characteristic directional spread, asymmetric growth, and day-to-day variability consistent with the observed fire evolution. This highlights an important strength of the proposed approach: its ability to integrate environmental forcing and geometric structure in a way that remains robust even when full observational datasets are unavailable—an increasingly common limitation in operational and post-event wildfire analyses.

For this example, we selected several representative locations across the fire-affected region and estimated the head-fire and back-fire rates of spread, along with wind speed and direction at those points. These estimated quantities were then used to construct the corresponding generalized elliptical frames that characterize the local fire-spread geometry. The enclosed area produced by these generalized ellipses provides a coherent approximation of the evolving burned region, even in the absence of complete ground-based measurements.

This  illustrative examples demonstrate the potential applicability of the proposed framework for modeling wildfire propagation using directional rates of spread. Although a full proof of concept would require extensive multi-temporal satellite datasets, which lie beyond the scope of this work, the examples highlight the strength and flexibility of the method in capturing realistic anisotropic fire-front dynamics from synthetic directional inputs.

\section*{Conclusion and Final Remarks}

This study presented a conceptual and methodological framework for modeling wildfire propagation using satellite-derived datasets integrated with mathematical modeling techniques. The proposed approach represents an initial step toward a comprehensive, data-driven fire spread model capable of capturing directional and terrain-dependent anisotropies in fire behavior without relying on detailed fuel or topographic measurements.

The hybrid framework combines thermal, atmospheric, and vegetation datasets within a Huygens-based geometric in whihc generalized elliptical frames are used to describe the evolution of fire fronts as expanding anisotropic wavefronts. In parallel, an alternative geometric strategy is incorporated in which the burned area is obtained by enclosing the region defined by generalized elliptical frames. By integrating thermal observations with vegetation indices and meteorological fields, the overall methodology infers spatially variable rates of spread that reflect the influence of live fuel moisture and wind. Although quantitative validation with real fire-event data is beyond the scope of this initial study, this omission is intentional: the emphasis here is on establishing the mathematical and computational foundations required for future data-driven evaluation.

We provided two examples to   illustrate how the framework combines  the satellite data with geometric techniques, even in data-limited scenarios. Using satellite-derived approximations of wind, vegetation conditions, and fire-front geometry, we demonstrated that both the Huygens-based method and the generalized elliptical-frame approach can reconstruct plausible fire-spread dynamics, underscoring the practical flexibility of the model.

It is important to emphasize that satellite observations alone, despite their global coverage, cannot deliver complete fire spread models due to limitations such as coarse spatial resolution, temporal gaps between overpasses, and cloud interference. These challenges reinforce the need to combine remote sensing information with mathematical and physical formulations, as proposed here.

A key strength of this methodology lies in its conceptual clarity, computational efficiency, and practical applicability. It bridges the gap between advanced mathematical modeling and satellite-based monitoring, offering a physically interpretable and scalable framework that can be implemented without heavy computational resources.

In summary, this work lays the groundwork for future studies aimed at calibrating and validating the model using real satellite and meteorological datasets. The proposed approach demonstrates how integrating multi-source satellite data with wavefront-based mathematical techniques can advance wildfire propagation modeling toward practical, scalable, and physically consistent solutions. Owing to its general structure and adaptability, the framework is suitable for both regional and global fire-propagation studies, supporting applications ranging from local fire management to large-scale environmental monitoring.

\newpage

\appendix
\section{Steps to predict the rates of fire spread}\label{apen}

The proposed approach to predict the rates of fire spread follows an iterative, data-driven procedure based on satellite imagery and    vegetation indices. The above methodology is listed as steps below:

\begin{enumerate}
    \item \textbf{Input data acquisition.} Obtain two successive satellite images capturing the progression of a wildfire starting from a point ignition source.

    \item \textbf{Fire front extraction.} From the images, extract the positions of the fire fronts and compute the initial estimates of the head fire and backfire rates of spread, denoted $R_H^T$ and $R_B^T$.

    \item \textbf{LFMC and wind sampling.} For several points along the extracted fire fronts, collect the corresponding LFMC and wind values from available vegetation index datasets.

    \item \textbf{Rate of spread refinement.} Use numerical fitting techniques to refine the initial spread rates, resulting in $R_H$ and $R_B$, the refined head and backfire spread rates.

    \item \textbf{Wavefront propagation.} Substitute the refined rates $R_H$ and $R_B$, together with the wind direction and velocity into the generalized elliptical fire spread model (Eq.~\eqref{ind}). Apply Huygens' principle to compute the next fire front or estimate generalized elliptical frames to approximate the perimeter of the burnt area.

    \item \textbf{Iterative update.} On the newly generated fire front, repeat Steps 3–5:
    \begin{itemize}
        \item Sample LFMC and wind values at selected points near the new front.
        \item Recalculate $R_H$ and $R_B$ using the updated LFMC and wind.
        \item Propagate the next wavefront using the updated elliptical model.
    \end{itemize}

    \item \textbf{Output generation.} After several iterations, obtain a sequence of predicted fire fronts modeling the spatial and temporal evolution of the wildfire under vegetation-dependent conditions.
\end{enumerate}

\bibliography{refs}
\bibliographystyle{unsrt}

\end{document}